# Observation of 1D Fermi arc states in Weyl semimetal TaAs

Xiaohu Zheng[1,2,†], Qiangqiang Gu[1,†], Yiyuan Liu[1], Bingbing Tong[1,2], Jian-Feng Zhang[1], Chi Zhang[3], Shuang Jia[1], Ji Feng[1,4,5,*], and Rui-Rui Du[1,4,5,*]

[1]International Center for Quantum Materials, School of Physics, Peking University, Beijing 100871, China

[2]Beijing Academy of Quantum Information Sciences, Beijing 100193, China

[3]State Key Laboratory of Superlattices and Microstructures, Institute of Semiconductors, Chinese Academy of Sciences, Beijing 100083, China

[4]CAS Center for Excellence in Topological Quantum Computation, University of Chinese Academy of Sciences, Beijing 100190, China

[5]Collaborative Innovation Center of Quantum Matter, Beijing 100871, China

† These authors contributed equally to this work

*Correspondence to: rrd@pku.edu.cn; jfeng11@pku.edu.cn

**Abstract**

Fermi arcs on Weyl semimetals exhibit many exotic quantum phenomena. Usually considered on atomically-flat surfaces with approximate translation symmetry, Fermi arcs are rooted in peculiar topology of bulk Bloch bands of three-dimensional (3D) crystals. The fundamental question of whether a 1D Fermi arc can be probed remains unanswered. Such answer could significantly broaden potential applications of Weyl semimetals. Here, we report a direct observation of robust edge states on atomic-scale ledges in TaAs using low-temperature scanning tunneling microscopy/spectroscopy. Spectroscopic signatures and theoretical calculations reveal that the 1D Fermi arcs arise from the chiral Weyl points of bulk crystal. The crossover from 2D Fermi arcs to eventual complete localization on 1D edges was arrested experimentally on a sequence of surfaces. Our results demonstrate extreme robustness of the bulk-boundary correspondence, which offers topological protection for Fermi arcs, even in cases in which the boundaries are down to atomic-scale. The persistent 1D Fermi arcs can be profitably exploited in miniaturized quantum devices.

**MAIN TEXT**

**Introduction**

Exploring the exotic properties of quasiparticles in topological matters is of great interest in condensed-matter physics[1–3]. Weyl semimetals have been theoretically proposed[4–6] and experimentally confirmed[7–13] as an important gapless topological system, which harbors pairs of Weyl points with opposite chirality. As a periodic cross section of the Brillouin zone (BZ) moves across a Weyl point, the Chern number changes from 0 to 1, and back to 0 as it continues across the other Weyl point of the opposite topological charge[6,14]. Consequently, the 2D cross sections between a pair of Weyl points can be viewed as a continuous *k*-space stack of Chern insulators, leaving a streak of chiral edge states on the surface, forming Fermi arc states[5,15]. In a number of natural gapless crystals, Fermi arc surface states have been identified on surfaces with connecting chiral Weyl points[4–12,16–21]. The (001) face in TaAs is a representative example of what is termed an arc-allowed surface (AAS)[6–8,13,20,21], as schematically shown in Fig. 1A. On the other hand, on an achiral surface on which the projections in the 2D surface BZ of the Weyl points coincide, topological Fermi arcs are not expected to exist[6], and these are referred to as an arc-forbidden surface (AFS). TaAs surfaces with Miller indices (100) and (110) are AFSs; whereas, (112) and (114) surfaces (see Fig. 1A), departing from the (110) AFS, can host the projection of chiral Weyl points. These surfaces have not yet been experimentally investigated from a Weyl physics perspective.

Evidently, the Fermi arcs of a Weyl semimetal, and their presence or absence, are derived from the peculiar topology of the Bloch bands of the bulk crystal, which inherently assumes an ideal, infinite crystalline system[14]. Although, in practice, one expects the Bloch band picture to hold for finite crystals with finite surfaces comprised of a large number of unit cells, the fate of Fermi arcs in structures down to atomic-scale, such as a step ledge, has yet to be examined experimentally. Here, we report a direct observation of one-dimensional (1D) edge states associated with Fermi arcs residing at the step edges on an AFS of a TaAs crystal, as well as on AASs with weak Fermi arc surface states. These edge states can be viewed as topological Fermi arcs that survive persistently on atomic-scale 1D step ledges, at which the Bloch theorem is not expected to apply. Spectroscopic signatures from a sequence of surfaces gradually departing from AAS (001) to approach AFS (110) show that the Fermi arcs undergo a continuous crossover from 2D surface states, to eventually complete localization on 1D step edges. Our results indicate that the bulk-boundary correspondence that protects the Fermi arc states is more ubiquitous than recognized previously. Indeed, these topologically-protected states exist not only on 2D surfaces, but also on 1D step edges, the latter of which can be used to create interesting 1D quantum devices or a Weyl semimetal single crystal with contiguously-covered topological surface states.

**Results and discussion**

A sequence of atomically-flat surfaces with different Miller indices in high-quality TaAs single crystals was measured comprehensively at 4.2 K in a commercial STM system (UNISOKU-1300) (see Materials and Methods). Tunneling spectra and differential conductance (dI/dV) mappings reveal uniform 1D edge states at the step edge of the (110) and the (112) surfaces. However, no signature of such edge states was observed at the steps on the (001) and the (114) surfaces. The particular spatial and energy distributions can distinguish these unique states from common trivial edge states originating from defects, such as dangling bonds. The correspondence between the experimental results and the theoretical calculations show that the observed edge states originate from the localization of Fermi arcs at the step edges.

**Discovery of 1D edge states on cleaved (112) surface**

We started from an atomically-flat, pristine (112) surface prepared by *in-situ* cleaving. A 3D topographic STM image, exhibiting long and straight step edges, is shown in Fig. 1B. Each terrace step is found to be one-atom high, i.e., approximately 2.4 Å. Fig. 1C shows an atomically-resolved topographic image of the top terrace, which agrees precisely with the configuration of (112). The step edge is along the $[1\bar{3}1]$ direction, which is the intersection between the (112) and the $(\bar{1}14)$ crystal facets, and has an inclination angle of approximately 52° from $[\bar{1}10]$ (right panel in Fig. 1C). Tunneling spectroscopy was performed on selected points near a step edge, as marked in Fig. 1B. As shown in Fig. 1D, a clear peak near the Fermi level in the spectra appears proximal to the step edge, in comparison with the spectra that are far from the line step (local density of state (LDOS) on the 2D surface), suggesting localized electronic states near the step edge. In the tunneling conductance mapping presented in Fig. 1E, data were collected on the same area as shown in Fig. 1B with various bias voltages. Trivial dangling bonds states can be observed, although mainly confined on the edge atoms around -200 meV and 250 meV. Of special interest are those uniform states next to the step edges at the energy of 55 meV, which correspond to near-zero energy peaks, as indicated by the dotted green line in Fig. 1D. Distinguished from the trivial edge states, these localized edge states disperse from the step to the surface with a width over ~ 1 nm in real space. Further analyses will demonstrate that these are, in fact, remnants of Fermi arcs, although these measurements are taken on an AFS. The above findings are highly interesting, since extant theoretical works have predicted that Fermi arcs can appear as localized states at 1D step edges in a 3D Weyl system[22,23]. We will focus on spectroscopic evidence for these 1D Femi arc states in this paper.

**1D edge states originate from topological Fermi arcs**

To elucidate the origin of the 1D edge states and their possible connection to 2D Fermi arcs, we systematically investigate surface states on a sequence of planes $(1, 1, 2n)$, of which $n = 0$ is AFS (110), $n = \infty$ is AAS (001), and $n = 1, 2$ denotes the nearest surfaces that depart from (110), i.e., (112) and (114), as shown in Fig. 1A. The STM topographic image captured on the surfaces reveal a number of atomically-flat facets. By comparing it with the structural models shown in Figs. 2A-C, a number of highly-crystalline facets are identified to be (110), (112) and (114) planes, all containing [$\bar{1}$10] step edges, as shown in Figs. 2D-F. The [$\bar{1}$10] step edge is parallel to the (001) AAS surface, and has a height of several atoms. In the following, we will examine the signal of Fermi arcs on these facets.

Fig. 3B shows the selected tunneling conductance (dI/dV) spectra along the arrow-dotted line in Fig. 3A on the AFS (110) facet as the tip approaches the [$\bar{1}$10] step edge. Non-zero LDOS at the Fermi level indicates the (semi-) metallic nature of the surface. The tunneling spectrum taken far away from the step, representing the 2D surface states, exhibits a small dip near the Fermi level. As the probe was moved toward the step edge, the dip in the tunneling spectrum gradually disappears while a peak grows steadily near the Fermi level.

Fig. 3C shows the spatial STS along the dashed arrow line in Fig. 3A. The red line indicates the position and the profile of the step edge. Prominent STS features corresponding to the peaked DOS are observed next to the step edge with a dispersion width in real space. STS mappings with selected bias voltages are presented in Fig. 3D. The trivial edge states that originate from the dangling bonds can be observed at -80 meV, which are precisely confined on the edge atoms. The peaked edge states at -20 meV (peaked DOS in STS), however, spread over a whole unit cell along the $\vec{c}$ direction into the surface. The dispersion width corresponds perfectly to the dispersion depth of Fermi arc surface states on AAS (001) [13](Fig. S1), and can be considered as the profile of the Fermi arc surface states of an atomic (001) ledge. This result further suggests the topological origin of the edge states[24,25].

To understand the origin of the emergent peak in STS spectra near the step edge, electronic structure calculations were carried out to describe the low-energy excitations of the step edge in question. A slab model with terraced surfaces on the top and bottom was used, with Miller indices $(n, n, 2)$ (Materials and Methods, and Supplementary Materials). The flat region on the terrace has a width of $n|\vec{c}|$, with the (001) plane (As-terminated for top and Ta-terminated for bottom) exposed at the step ledges with a height of unit cell $|\vec{a} + \vec{b}|$. The Ta-As chains propagate along the $[\bar{1}10]$ direction (cf. Fig. 2A). Based on a tight-binding Hamiltonian obtained from density-functional theoretic calculations, surface Green's functions were obtained with an iterative technique to yield surface spectral function for direct comparison with the tunneling spectra[26].

For the model described above, the flat region is the (110) AFS, and consequently one expects to see topological states only near the step edges, which corresponds precisely to the experimental finding. The spectral functions $A(k, \varepsilon)$ on $(n, n, 2)$ ($n = 6$ in Fig. 3E) for $\varepsilon = -20$ meV on the top surface are displayed in Fig. 3F. It can be seen that both Fermi arcs and trivial Fermi surfaces are present in the surface BZ. Moreover, the Fermi arcs can be observed in the energy range between -5 meV to -30 meV with their maximum intensity at -20 meV, which corresponds to the peaked DOS in the spectra signatures (see Fig. S6). To determine the spatial location of the Fermi arcs on the terraced surface with a width over 7 nm (12 Ta-As chains), we chose one of the Fermi arcs that is clearly isolated from other surface states, and computed its projection on different Ta-As chains as labeled in Fig. 3E on the terraced slab model, which are depicted in Figs. 3F and G. Other visible arcs on the top and the bottom surfaces at $\varepsilon = -20$ meV are shown in Fig. S5. It is evident that the spectral weight of the Fermi arcs is most pronounced at the step edge, and decreases steadily as the distance from the step increases (Figs. 3G and H). The calculation results support the existence of the 1D Fermi arc edge states at the step edge on the (110) surface, which disperse from the edge to the surface with a width of ~ 1 nm in real space.

**Crossover from 2D Fermi arc states to 1D Fermi arc edge states**

The atomically-resolved STM topography of (112) in Fig. 2E shows relaxation of surface atoms induced by the annealing process (Fig. S3). Fig. 4A presents the STM topographic surface plates (112) with steps. The height profile across the surface shows that the height of the small step ledge is approximately 0.85 nm (four atomic-layers). In addition, the small ledge is along $[\bar{1}10]$ and terminates on the (001) surface, which permits the existence of Fermi arc surface states. For the tunneling spectra acquired far from the step edge, a small shoulder is observed in the curves at a bias energy of -20 meV. When the STM tip approaches the step edge, the shoulder evolves into a pronounced peak in the *dI/dV* spectra in the energy range between -20 meV to 20 meV (Fig. 4B). The near-zero energy peaks are distinguishable from the trivial edge states which are precisely confined on the edge atoms (Fig. 4B). They also spread over several atoms on both sides of the step (Fig. 4B), and even disperse across the narrow terrace R2 (~ 1.3 nm) (Fig. 4B). The 1D edge states revealed by the near-zero energy peaks can be clearly seen in the spatial revolved *dI/dV* along the step edges (Fig. 4C). The peak positions near the steps are spatially steady and essentially energy-independent (i.e., without interference patterns), as shown in Fig 4C. The lack of interference rules out the possibility that the standing wave originated from the scattered electrons on the steps[24].

A slab model with terraced $(n, n, 2n+1)$ surfaces on the top (As-terminated) and bottom (Ta-terminated) surfaces was also constructed to illustrate the 1D edge states on the (112) terrace. The level region on the terrace has a width of $n|\vec{c} - \vec{a} - \vec{b}|$ ($n = 4$, width ~5 nm), with the (001) surface exposed at the atomic-thick step ledge with a height of unit cell $|\vec{b}|$, as schematically illustrated in Fig. 4D. The spectral functions $A(\vec{k}, \varepsilon)$ of the terraced surface for $\varepsilon = -20$ meV on the top surface (bottom surface in Fig. S8) are displayed in Fig. 4E. The visible Fermi arcs were selected (in Fig. 4E) to calculate the projection weight on Ta-As chains as labeled in Fig. 4D. The spectral weight exhibits pronounced localization at the step edge (Fig. 4F), which accounts for the observed 1D edge states in Figs. 4B and C. Noticeably, it decreases more slowly than that on the (110) terrace as the distance from the step increases, indicating the existence of 2D topological Fermi arc surface states.

Remarkably, the 1D Fermi arc edge states appear to coexist with the 2D Fermi arc surface states in this case.

The Fermi arc states in Weyl semimetals, protected by the peculiar topology of the Bloch bands of the bulk crystal, are robust against weak surface perturbations[14,20,27]. Here, we examine how the near-zero energy edge states respond to local perturbation. In Fig. 4A, the whole region is divided into three areas, R1, R2 and R3, of which their width is R1 (~8 nm) > R3(~2 nm) > R2(~1.3 nm). The metallicity (surface DOS intensity) of each area revealed by spectroscopic signatures follows the relationship: R1 > R3 > R2 (Fig. 4G). The peak in the 1D edge states is robust and only scarcely affected when the change of the surface size and metallicity are considered as weak perturbations. Moreover, the peak also shows protection against the weak disorder of local defects. Two kinks induced by the edge parallel translation can be discerned, as shown in Fig. 4A. We took *dI/dV* curves spatially at each point as numbered from No. 1 to 22 with minimal separation distance (~0.15 nm) along the step. The left panel in Fig. 4H presents the possible atomic configuration of the edge in accordance with that in Fig. 4A. Configuration of the dangling bonds in the kinks is different (positions No. 5-8 and No. 16-19), in the sense that the kinks can be considered as disorders or point defects to disturb the LDOS. If the 1D edge states were of trivial origins, the corresponding peaks should be changed by local defects. However, no substantial changes in the spatial revolved *dI/dV* spectra at each numbered point (right panel in Fig. H) is found, which lends further support to its topological nature.

The surface plane of (114) with a step edge along [$\bar{1}$10] (Fig. 5A) is the last member of the sequence that we prepared. The (114) surface exhibits a decreased inclination angle (~50º) in respect to the AAS (001). The surface states on (114) have been detected. Although the LDOS at Fermi level increases slightly as the STM tip approaches the step edge, no extra-peaked STS features can be seen near the step edge (Fig. 5B). In the spatial spectra results (Fig. 5C), the dangling bond states can be observed on the Ta-As chains at energy above the Fermi level, and increased DOS are also seen on the inclined ledge. However, no uniform 1D edge states with possible topological origin can be discerned. Our theoretical calculations on (114) demonstrate that Fermi arcs distribute all over the surface, and the projections of Fermi arcs on all Ta-As chains are of comparable weight (Figs. 5E-G). This suggests that the 2D Fermi arc surface states dominate the topological information on the (114) surface, which is similar to the (001) surface with steps where no topological edge states exist owing to the 2D Fermi arc all over the surface (Fig. S2). This confirms the expectation that, as the surface becomes closer to AAS (001), the spectroscopic signature of the surface states increasingly resembles that obtained on the (001) surface. The above results also verify that, when the surface indices are gradually departing from AFS (110) and approaching AAS (001), the Fermi arcs undergo a continuous crossover from 1D edge states to completely 2D topological surface states.

In Fig. 1, it can be seen that plane (112) has the step edge along $[1\bar{3}1]$. Since $[1\bar{3}1]$ can be viewed as a ledge of AAS ($\bar{1}14$) where Fermi arc surface states exist, the chiral Weyl points have a finite weight projected on the ledge, and consequently the 1D Fermi arc edge states appear at the step edge. In aggregate, the results further confirm that the 1D Fermi arc states exist ubiquitously in 3D Weyl crystal step edges.

**Conclusions**

In this work, we not only observed 1D Fermi arc edge states, but also explored the evolution of Fermi arc states. In TaAs crystal, the (110) and (001) facets are perpendicular to each other. When a step is formed by the (110) surface (AFS) and the (001) ledge (a finite AAS), the Fermi arcs can only appear on the AAS ledge with a certain penetration depth on the AFS, which therefore forms localized Fermi arc states on the edge. As the surface index changes, however, an evolution of the Fermi arc states can be observed from our calculations. To elucidate the evolution of Fermi arc states, we examined the steps' form by the $(1, 1, 2n)$ surface with the (001) ledge. For $n = 0$, the surface is the AFS (110), and for $n = \infty$ the surface is the AAS (001). As shown in Fig. 4 where $n = 1$, the step is formed by two AASs (112) and (001), and the Fermi arcs can survive on both facets. Therefore, on the (112) facet of the step, the coexistence of 2D Fermi arc surface states and 1D Fermi arc edge states can be observed simultaneously in the calculated results in Fig. 4F. By increasing the index $n$, the surface $(1, 1, 2n)$ approaches the surface (001), and the 2D Fermi arc surface states on the $(1, 1, 2n)$ facet become more prominent. In the case of $n = 2$ (Fig. 5), the step is formed by two AASs (114) and (001), and the 2D Fermi arc surface states dominate on the top (114) facet.

Overall, the STM/STS measurements and the theoretical calculations performed in this work demonstrate that 1D topological Fermi arc states widely exist on atomic step edges, which can be conceptually viewed as the projection of chiral Weyl points in the bulk of Weyl semimetal TaAs. In addition, the 1D Fermi arc edge states undergo a continuous crossover to the 2D surface states as the surface gradually departs from an AFS and approaches an AAS, and both the 1D and the 2D Fermi arc states may coexist in the process. The results reveal that the bulk-boundary correspondence in 3D Weyl semimetals remains at work even when the boundary is down to the atomic-scale.

**Materials and Methods**

**Details of the sample preparation:** High-quality single crystals of TaAs were grown by the standard chemical vapor transport method, as described in [28]. For the processed TaAs samples: the surface was polished by abrasive papers after the (110) surface was demarcated by Laue diffraction. Then, the sample was transferred into an ultra-high vacuum chamber and repeatedly sputtered by $Ar^+$ ions with energy 500 eV. The annealing process was carried out on the sample by electron beam heating with a temperature of approximately 950 °C for 30 min under a vacuum of $10^{-9}$ Torr. For the cleaved samples: the thickness of synthetic TaAs crystals ((001) plane) was polished down to approximately 300 μm with abrasive papers from both sides. It was fixed to the sample holder for cleavage on the (110) plane. Cleavage of the sample was carried out *in-situ* in a high vacuum chamber ($2.5 \times 10^{-10}$ Torr) at room temperature, with a cleaving knife equipped in the Unisoku-1300 STM/STS system. After the cleavage, the sample was transferred without interrupting the high vacuum into the STM chamber. After numerous rounds of trial and error, a region of the (112) plane was captured by STM measurement.

**STM/STS measurements:** STM/STS are performed at liquid helium temperature (4.2 K) in the Unisoku-1300 system with a Nanonis controller and the built-in lock-in amplifier. Tungsten tips were used in all of the STM/STS measurements. In the measurement of the topographic images, the constant current mode was used with the setting sample bias $V_{bias}$= 100 mV and $I_{setpoint}$= 500 pA. When performing the tunneling spectra (the *dI/dV* curves) and conduction maps, lock-in techniques were used with a modulation amplitude of 3-5 mV, frequency of 707 Hz, $V_{bias}$=200 mV, and $I_{setpoint}$=500 pA to 1 nA. The difference of energy interval that arises on prominent 1D edge states on cleaved (112) and processed (112) in Fig. 1 and Fig. 4, respectively, may be ascribed to the details of doping in different batches of samples.

**Calculations:** The *ab initio* calculations were performed using the Vienna *ab initio* simulation package (VASP)[29] within the generalized gradient approximation (GGA) parametrized by Perdew, Burke, and Ernzerhof (PBE)[30]. The Kohn–Sham single-particle wave functions were expanded in the plane wave basis set with a kinetic energy truncation at 400 eV. The crystal structure of the unit cell of TaAs was fully relaxed until Hellmann–Feynman forces on each atom were less than 0.001 eV/Å with a 12×12×3 k-mesh sampled in the BZ. To calculate the surface and bulk electronic structure, a tight-binding Hamiltonian was constructed using the VASP2WANNIER90 interface[31]. The surface states' electronic structures were calculated by the surface Green's function technique[26].

**Acknowledgments**

**Funding:** This work was supported by the National Key Research and Development Program of China (Grants No. 2016YFA0301004, 2017YFA0303301, and 2018YFA0305601); by the Strategic Priority Research Program of the Chinese Academy of Sciences (Grant No. XDB28000000); by the National Natural Science Foundation of China (Grants No. 11725415, 11921005, and 11934001); and by the NSFC Young Scientists Program (Grant No. 11704010).

**Author contributions:** R. D., J. F., X. Z., and Q. G. devised the experiments and calculations; X. Z. performed the STM/STS measurements; Q. G. performed theoretical calculations; S. J. and Y. L. synthetized the TaAs crystals; R. D., J. F., X. Z., and Q. G. analyzed the data, and wrote the paper with input from the other co-authors.

**Conflict of interest statement**

None declared

**Data and materials availability:** All data needed to evaluate the conclusions in the paper are present in the paper and/or the Supplementary Materials. Additional data related to this paper may be requested from the authors.

**References**

1. Bansil A, Lin H, Das T. *Colloquium* : Topological band theory. *Rev. Mod. Phys.* 2016; **88**:021004.

2. Burkov AA. Topological semimetals. *Nat. Mater.* 2016;**15**:1145–8.

3. Chiu C-K, Teo JCY, Schnyder AP *et al.* Classification of topological quantum matter with symmetries. *Rev. Mod. Phys.* 2016;**88**:035005.

4. Wan X, Turner AM, Vishwanath A *et al.* Topological semimetal and Fermi-arc surface states in the electronic structure of pyrochlore iridates. *Phys. Rev. B* 2011;**83**:205101.

5. Burkov AA, Balents L. Weyl semimetal in a topological insulator multilayer. *Phys. Rev. Lett.* 2011;**107**:127205.

6. Weng H, Fang C, Fang Z *et al.* Weyl semimetal phase in noncentrosymmetric transition-metal monophosphides. *Phys. Rev. X* 2015;**5**:011029.

7. Xu S-Y, Belopolski I, Alidoust N *et al.* Discovery of a Weyl fermion semimetal and topological Fermi arcs. *Science* 2015;**349**:613–7.

8. Lv BQ, Xu N, Weng HM *et al.* Observation of Weyl nodes in TaAs. *Nat. Phys.* 2015;**11**:724–7.

9. Schröter NBM, Pei D, Vergniory MG *et al.* Chiral topological semimetal with multifold band crossings and long Fermi arcs. *Nat. Phys.* 2019;**15**:759–65.

10. Yang Y, Sun H, Xia J *et al.* Topological triply degenerate point with double Fermi arcs. *Nat. Phys.* 2019;**15**:645–9.

11. Morali N, Batabyal R, Nag PK *et al.* Fermi-arc diversity on surface terminations of the magnetic Weyl semimetal $Co_3Sn_2S_2$. *Science* 2019;**365**:1286–91.

12. Liu DF, Liang AJ, Liu EK *et al.* Magnetic Weyl semimetal phase in a Kagomé crystal. *Science* 2019;**365**:1282–5.

13. Inoue H, Gyenis A, Wang Z *et al.* Quasiparticle interference of the Fermi arcs and surface-bulk connectivity of a Weyl semimetal. *Science* 2016;**351**:1184–7.

14. Armitage NP, Mele EJ, Vishwanath A. Weyl and Dirac semimetals in three-dimensional solids. *Rev. Mod. Phys.* 2018;**90**:015001.

15. Halász GB, Balents L. Time-reversal invariant realization of the Weyl semimetal phase. *Phys. Rev. B* 2012;**85**:035103.

16. Sanchez DS, Belopolski I, Cochran TA *et al.* Topological chiral crystals with helicoid-arc quantum states. *Nature* 2019;**567**:500–5.

17. Min C-H, Bentmann H, Neu JN *et al.* Orbital fingerprint of topological Fermi arcs in the Weyl semimetal TaP. *Phys. Rev. Lett.* 2019;**122**:116402.

18. Moll PJW, Nair NL, Helm T *et al.* Transport evidence for Fermi-arc-mediated chirality transfer in the Dirac semimetal $Cd_3As_2$. *Nature* 2016;**535**:266–70.

19. Huang S-M, Xu S-Y, Belopolski I *et al.* A Weyl Fermion semimetal with surface Fermi arcs in the transition metal monopnictide TaAs class. *Nat. Commun.* 2015;**6**:7373.

20. Sun Y, Wu S-C, Yan B. Topological surface states and Fermi arcs of the noncentrosymmetric Weyl semimetals TaAs, TaP, NbAs, and NbP. *Phys. Rev. B* 2015;**92**:115428.

21. Batabyal R, Morali N, Avraham N *et al.* Visualizing weakly bound surface Fermi arcs and their correspondence to bulk Weyl fermions. *Sci. Adv.* 2016;**2**:e1600709–e1600709.

22. Kodama K, Takane Y. Persistent current due to a screw dislocation in Weyl semimetals: role of one-dimensional chiral states. *J. Phys. Soc. Jpn.* 2019; **7**:054715.

23. Takane Y. Chiral surface states on the step edge in a Weyl semimetal. *J. Phys. Soc. Jpn.* 2017;**86**:104709.

24. Yang F, Miao L, Wang ZF *et al.* Spatial and energy distribution of topological edge states in single Bi(111) bilayer. *Phys. Rev. Lett.* 2012;**109**:016801.

25. Li Z, Zhuang J, Wang L *et al.* Realization of flat band with possible nontrivial topology in electronic Kagome lattice. *Sci. Adv.* 2018;**4**:eaau4511.

26. Sancho MPL, Sancho JML, Sancho JML *et al.* Highly convergent schemes for the calculation of bulk and surface Green functions. *J. Phys. F: Met. Phys.* 1985;**15**:851–8.

27. Wang Q, Zheng J, He Y *et al.* Robust edge photocurrent response on layered type II Weyl semimetal $WTe_2$. *Nat. Commun.* 2019;**10**:5736.

28. Van Mal HH, Buschow KHJ, Miedema AR. Hydrogen absorption of rare-earth (3d) transition intermetallic compounds. *J. Common Met.* 1976;**49**:473–5.

29. Kresse G, Furthmüller J. Efficiency of ab-initio total energy calculations for metals and semiconductors using a plane-wave basis set. *Comput. Mater. Sci.* 1996;**6**:15–50.

30. Perdew JP, Burke K, Ernzerhof M. Generalized gradient approximation made simple. *Phys. Rev. Lett.* 1996;**77**:3865–8.

31. Mostofi AA, Yates JR, Lee Y-S *et al.* wannier90: A tool for obtaining maximally-localised Wannier functions. *Comput. Phys. Commun.* 2008;**178**:685–99.

**Figure legends**

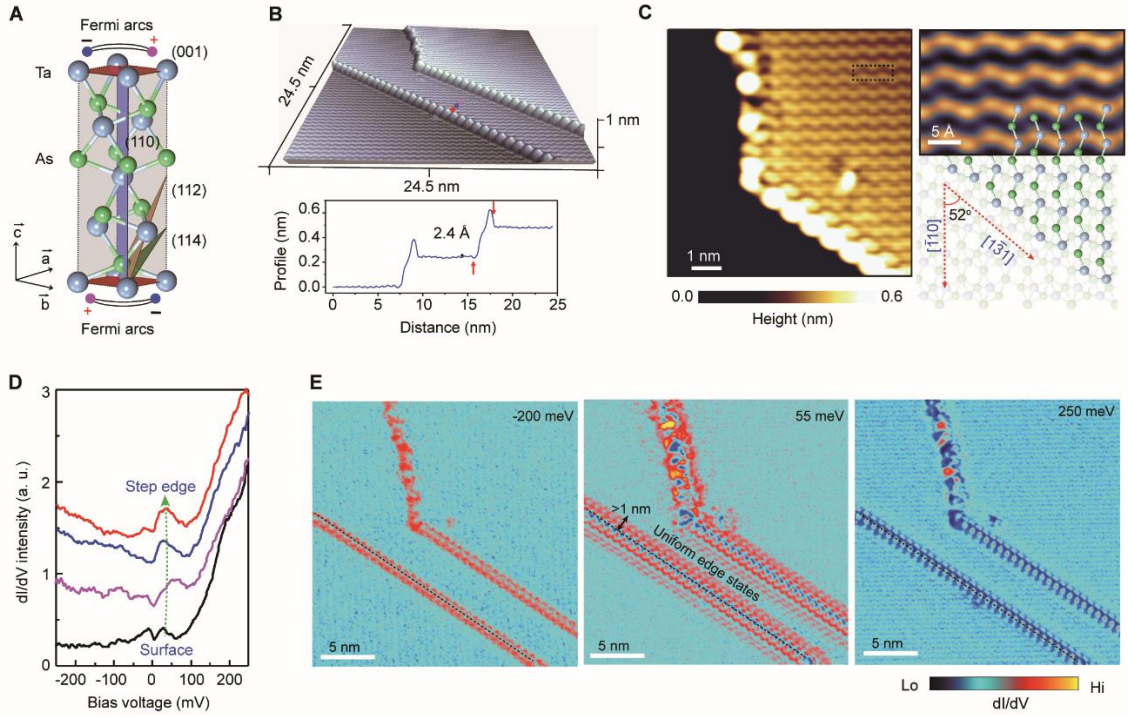

**Fig. 1. 1D edge states on the cleaved (112) surface.** (**A**) Schematic illustration of the unit cell of TaAs crystal where $\vec{a}, \vec{b}$, and $\vec{c}$ are the lattice vectors. The projection of chiral Weyl points and Fermi arcs is illustrated on (001) surfaces (top and bottom), and the crystal planes of (110) perpendicular to (001) host projection of chiral Weyl points coincide. As the Miller indices increase, the crystal facets depart from (110) to (001), such as the selected ones of (112) and (114); (**B**) 3D topographic STM image containing the atomic steps along $[1\bar{3}1]$ on the (112) surface, and the bottom panel shows that the step profile is one-atom thick; (**C**) High-resolution topographic image of the cleaved (112) surface with the step edge, in which the dashed square denotes a TaAs unit cell (left panel). The right panel shows a comparison between the surface configuration and the structural model. The step edge along $[1\bar{3}1]$ is an atomic ledge of the $(\bar{1}14)$ crystal facet, and has an inclination angle of ~52º from $[\bar{1}10]$; (**D**) *dI/dV* spectra captured near the step (labeled with color points in (B)) and 2D surface states far from the edge. A peak just above the Fermi level can be seen near the steps; (**E**) *dI/dV* mappings of the same region as presented in (B) with selected bias voltages show the trivial edge states right on the edge atoms at -200 meV and 250 meV, while the uniform edge states at 55 meV spread into the surfaces from both sides of the steps with a width over ~ 1 nm.

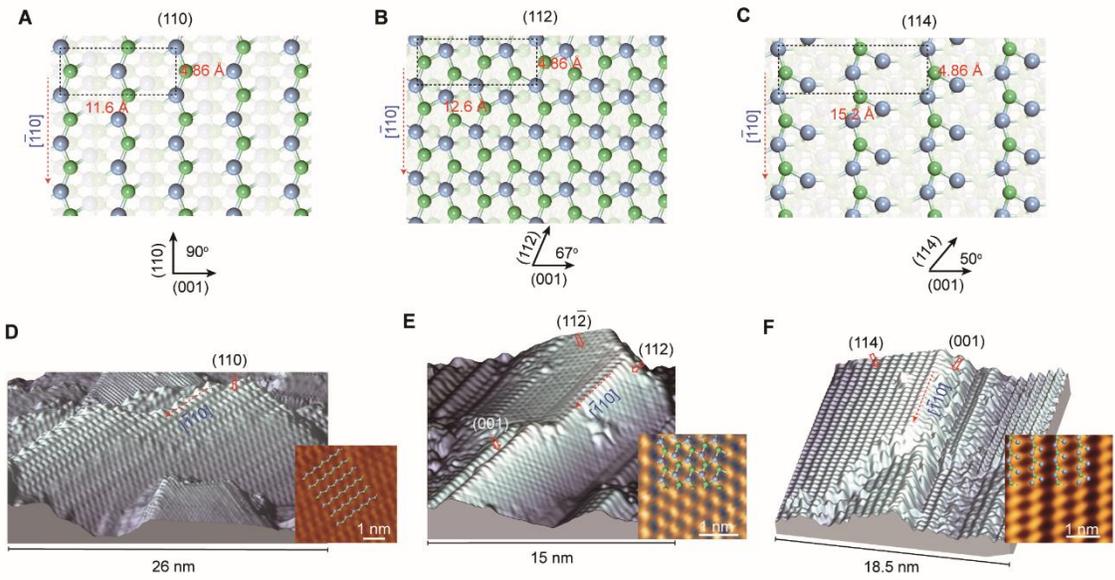

**Fig. 2. STM topographic images of (110), (112), and (114) planes in TaAs.** (**A**) to (**C**) Schematic illustration of the surface atomic structures of (110), (112) and (114) crystal planes, which have inclination angles of 90°, ~67°, and ~50° respectively, from the AAS (001), where each plane is formed by Ta-As chains along the [$\bar{1}$10] direction; (**D**) to (**F**) Atomically-resolved 3D STM topographic images of (110), (112), and (114) planes with their step edges along the [$\bar{1}$10] direction.

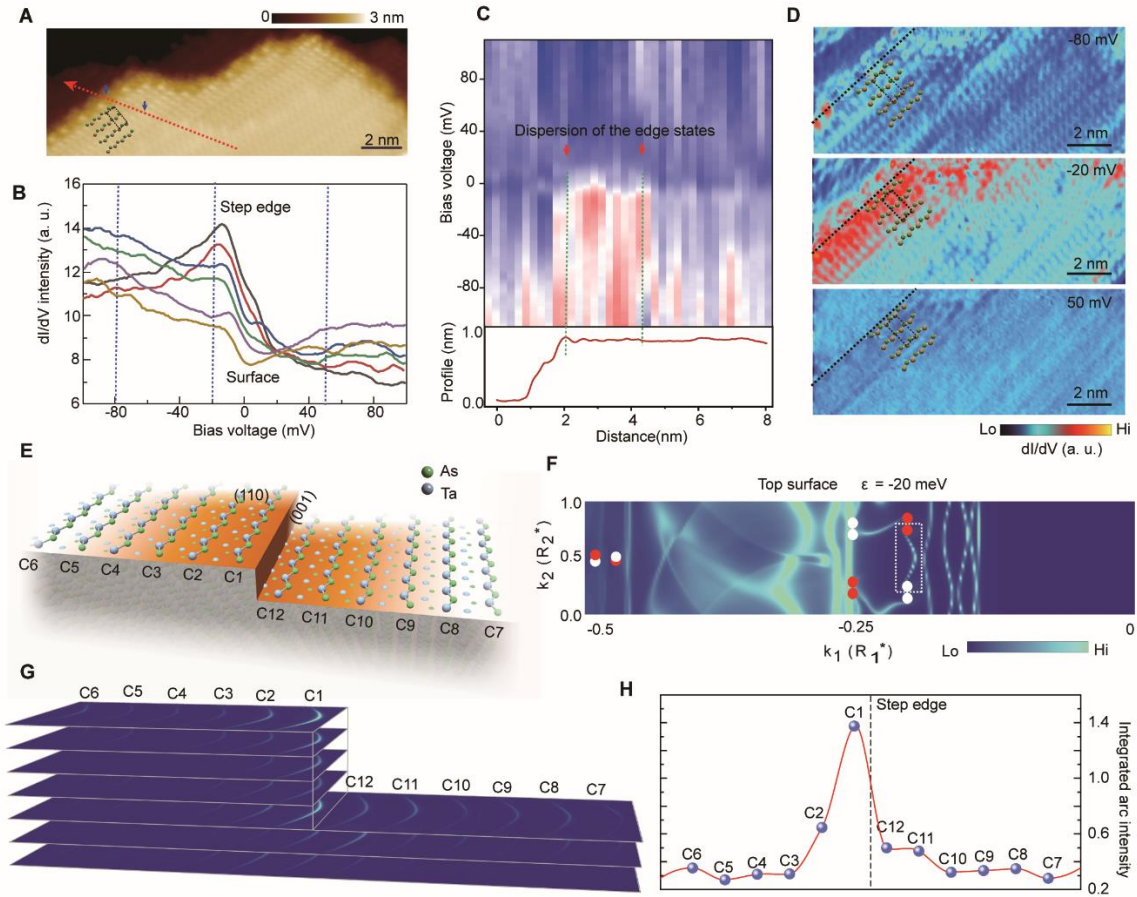

**Fig. 3. 1D edge states reside on the step edge of the (110) surface.** (**A**) The STM topographic surface plates (110) with step edges with the atomic structure model put on the surface; (**B**) The spatial-resolved *dI/dV* spectra along the arrow line as labeled in (A). Pronounced excited peaks in the tunneling spectra near the Fermi level are detected as the STM tip moves from the surface toward the step edge; (**C**) Up: the spatial STS, and down: the profile of the surface topography, along the dashed arrow line in (A), shows the dispersion width of the edge states next to the step; (**D**) *dI/dV* mappings taken on the step region in (A) at sample bias as marked in (B). Trivial edge states that originate from dangling bonds can be observed at -80 meV, pronounced and uniform 1D edge states can be seen at -20 meV corresponding to the peaked DOS in STS, and unique edge states can be observed spread over the unit cell next to the step; (**E**) Schematic illustration of part of the top surface in the calculation slab model with periodic steps jointing (110) and (001) planes; (**F**) Surface spectral functions $A(k, \varepsilon)$ in half the surface BZ at $\varepsilon = -20$ meV with respect to bulk Fermi energy. $A(k, \varepsilon)$ in the other half surface BZ can be inferred through time-reversal symmetry. $\vec{k}_1$ and $\vec{k}_2$ are reciprocal lattice vectors in the surface BZ defined by the lattice vectors $\vec{R}_1$ and $\vec{R}_2$, respectively, as detailed in Fig. S5. The selected Fermi arc is labeled by the dashed line, and the red and white dots denote the projection of chiral Weyl points; (**G**) Projection weight of the Fermi arc on Ta-As chains as numbered in **c** in momentum space; (**H**) Intensity of the integrated projection weight on Ta-As chains exhibits high localization at the step edge.

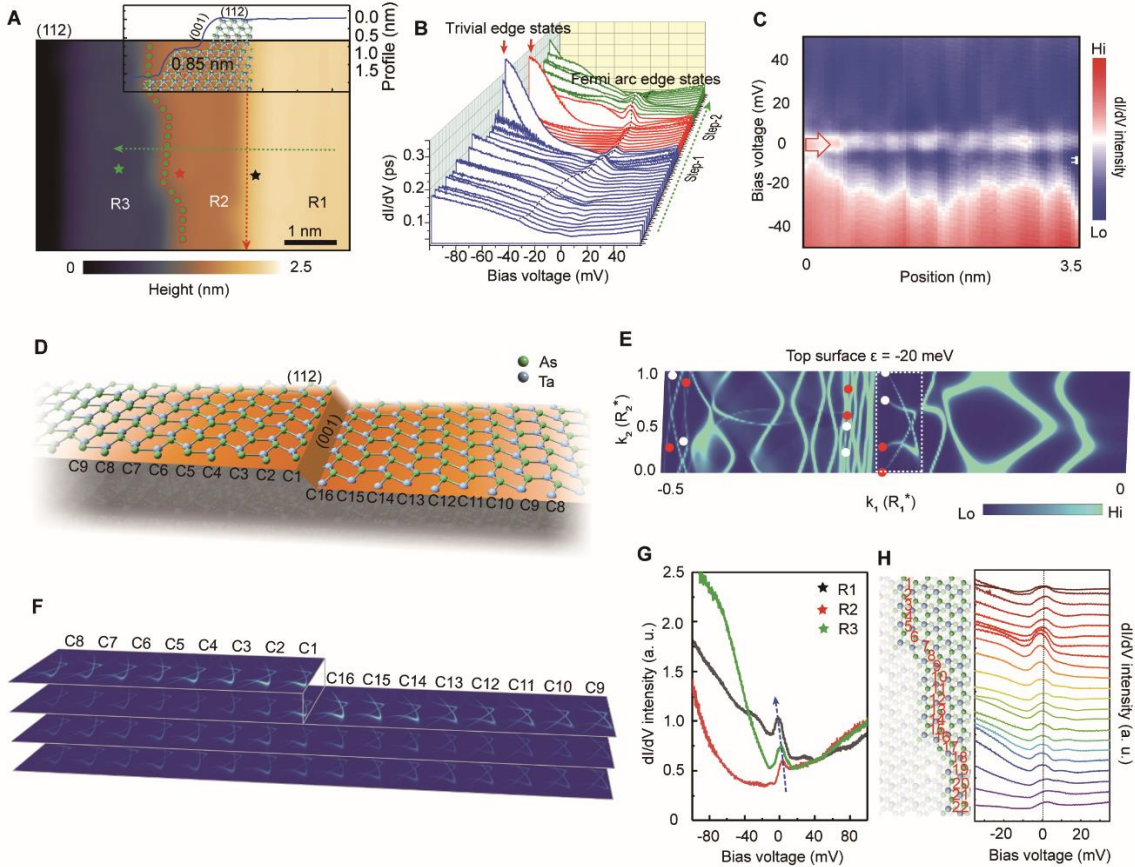

**Fig. 4. Topological states on the (112) plane.** (**A**) STM morphology of the (112) surface plane that contains steps dividing the plane into three terraces (R1, R2, and R3). The height profile across the surface is shown in the inset; (**B**) Spatial-resolved *dI/dV* spectra acquired in (A) along the green dashed arrow line. Peaks near the Fermi level become prominent with an energy shift when the STM tip approaches step-1, and peaks can be observed over the terrace of R2. The trivial edge states that originate from the dangling bonds emerge confined on the edge of both step-1 and step-2; (**C**) One-dimensional (1D) spectroscopic line along the [$\bar{1}$10] step (red dashed arrow in (A)). 1D edge states can be observed in the spectra; (**D**) Schematic illustration of the step jointing (112) and (001) planes on the top surface of the calculation slab model. The details of this model can be seen in Fig. S7. The (112) terrace is more than 5 nm width with 16 Ta-As chains (C1-C16); (**E**) The surface spectral functions $A(k,\varepsilon)$ of the terraced surface for $\varepsilon = -20$ meV in half the surface BZ; (**F**) Spectral weight of the selected Fermi arc in (E) projected on Ta-As chains as numbered in (D); (**G**) The peaked topological edge states in *dI/dV* spectra acquired from R1, R2, and R3 (star positions) show a small shift due to the different intensity of quantum confinement effect between terraces; (**H**) Left panel: schematic illustration of the step configuration in (A) with points (1 to 22) where *dI/dV* curves were measured. Two kinks can be observed along the step edge. Right panel: schematic illustration that shows that low-energy peaks in *dI/dV* spectra are robust without significant change, even at the kinks.

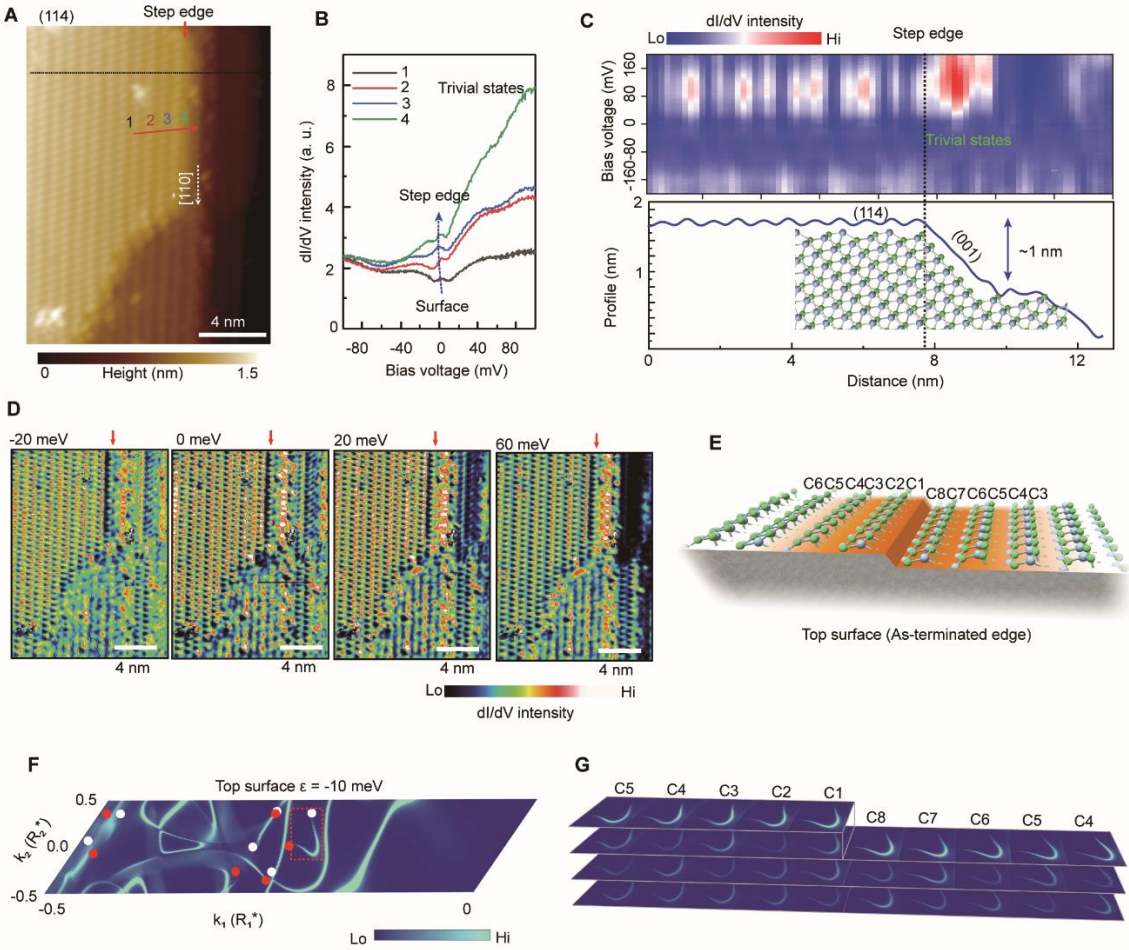

**Fig. 5. Surface and edge states on the (114) plane.** (**A**) STM morphology of the (114) plane with step edge along [$\bar{1}$10]; (**B**) Selected *dI/dV* spectra at position from 1 to 4, which gradually approach the step, show the evolution of LDOS. The whole LDOS is enhanced at the step edge, but no extra peaked DOS associated with the topological edge states can be seen; (**C**) Up: spatial tunneling spectra crosses the step edge. Down: the corresponding profile of the step where the ledge is formed by jointing (114) and (001) planes. The inset shows the corresponding atomic configuration of TaAs; (**D**) *dI/dV* conduction maps acquired at certain

sample bias from -40 to 60 mV. There is no indication that the step shows a localization of topological edge states, but above the Fermi level the increased DOS can be observed on the ledge of the step; (**E**) Periodic step edges jointing (114) and (001) facets are generated on the (229) surface with flat (114) terrace in the slab with As- terminated edge on the top surface. Details of the calculation model are presented in Fig. S9; (**F**) Surface Fermi surface (FS) plots at $\varepsilon = $ -10 meV on the surface in (E) show the surface states where Fermi arcs that connect chiral Weyl points can be observed. The other possible configurations for steps are calculated and presented in Fig. S10; (**G**) Projection of selected arcs can be observed on each of the Ta-As chains on the topmost surfaces without significant enhancement at the step edges.

# Supplementary Materials for

## Observation of 1D Fermi arc states in Weyl semimetal TaAs

Xiaohu Zheng[1,2†], Qiangqiang Gu[1†], Yiyuan Liu[1], Bingbing Tong[1,2], Jian-Feng Zhang[1], Chi Zhang[3], Shuang Jia[1], Ji Feng[1,4,5*], Rui-Rui Du[1,4,5*]

[1]International Center for Quantum Materials, School of Physics, Peking University, Beijing 100871, China.

[2]Beijing Academy of Quantum Information Sciences, Beijing 100193, China.

[3]State Key Laboratory of Superlattices and Microstructures, Institute of Semiconductors, Chinese Academy of Sciences, Beijing 100083, China.

[4]CAS Center for Excellence in Topological Quantum Computation, University of Chinese Academy of Sciences, Beijing 100190, China.

[5]Collaborative Innovation Center of Quantum Matter, Beijing 100871, China.

† These authors contributed equally to this work.

*Correspondence to: rrd@pku.edu.cn; jfeng11@pku.edu.cn.

**This PDF file includes:**

Fig. S1. Fermi arc states on (001) surface of TaAs.
Fig. S2. Step edge states on (001) surface of TaAs.
Fig. S3. Difference between processed and cleaved (112) surfaces.
Fig. S4. Construction of the calculation slab models with (110) terraced planes and atomic-thick (001) exposed at the step ledges.
Fig. S5. All visible Fermi arcs on the top and bottom surface with (110) terraces.
Fig. S6. Calculation of Fermi arc projection on (110) terraced surface as a function of energy.
Fig. S7. Construction of the calculation slab model with (112) terraced planes and atomic-thick (001) exposed at the step ledge.
Fig. S8. Surface states on the bottom surface with (112) terraces.
Fig. S9. Construction of the calculation slab model with (114) terraced planes and atomic-thick (001) exposed at the step ledge.
Fig. S10. 2D Fermi arc surface states on (114) terraced planes.

**Surface and edge states on Fermi arc-allowed-surface (001) in TaAs crystal**

The Fermi arc surface states distribute on the top and bottom surfaces of (001) crystal face in TaAs (as shown in Fig. S1A) owing to the broken inversion symmetry. Here, we have calculated the Fermi surfaces on As-terminated (001) facet, which contains the Fermi arcs, the trivial surface states and the bulk derived surface resonance states. By projecting the Fermi surface onto each of Ta- layers in the unit cell, we can see the trivial surface states decaying very fast, and survive prominently on the topmost surface. In contrast, the Fermi arcs can be observed on Ta- layers all over the unit cell. It demonstrates that Fermi arc surface states can disperse into the bulk with a particular depth in real space, which is comparable to the unit constant along $\vec{c}$ direction (~1.2 nm).

We performed the STM/STS measurements on the *in-situ* cleaved fresh (001) surface. The atomic resolved topographic image is presented in Fig. S1C, which contains high density point defects (vacancies). The defects induced the electrons scattering inter/intra the Fermi contours, and forming the quasiparticle interference (QPI) patterns on the STS mapping (acquired with the sample bias voltage of -25 mV), as shown in Fig. S1d. By Fourier transform, the scattering wave vector can be observed in the momentum space. Fermi arcs and their connection to the bulk continuum can be detected by the QPI patterns. Our results on (001) are in accordance with the previous works, which demonstrate the high quality of the TaAs crystals.

On the cleaved (001) surface, we investigated the states at the step edge. As shown in Fig. S2A, there is a step with a minimal height. The LDOS has been measured by STS, which shows no peaked DOS at the step edge, as shown in Fig. S2B. By carefully comparing the spectra at the surface with those near the step edge, it is clearly shown that the LDOS are perturbed locally at step edge (Fig. S2C). In the *dI/dV* mappings acquired at various of bias voltages near the Fermi level (Fig. S2D), interference patterns originate from the electronic standing waves on the surface and near the step can be observed, but again, no prominent and uniform edge states can be detected.

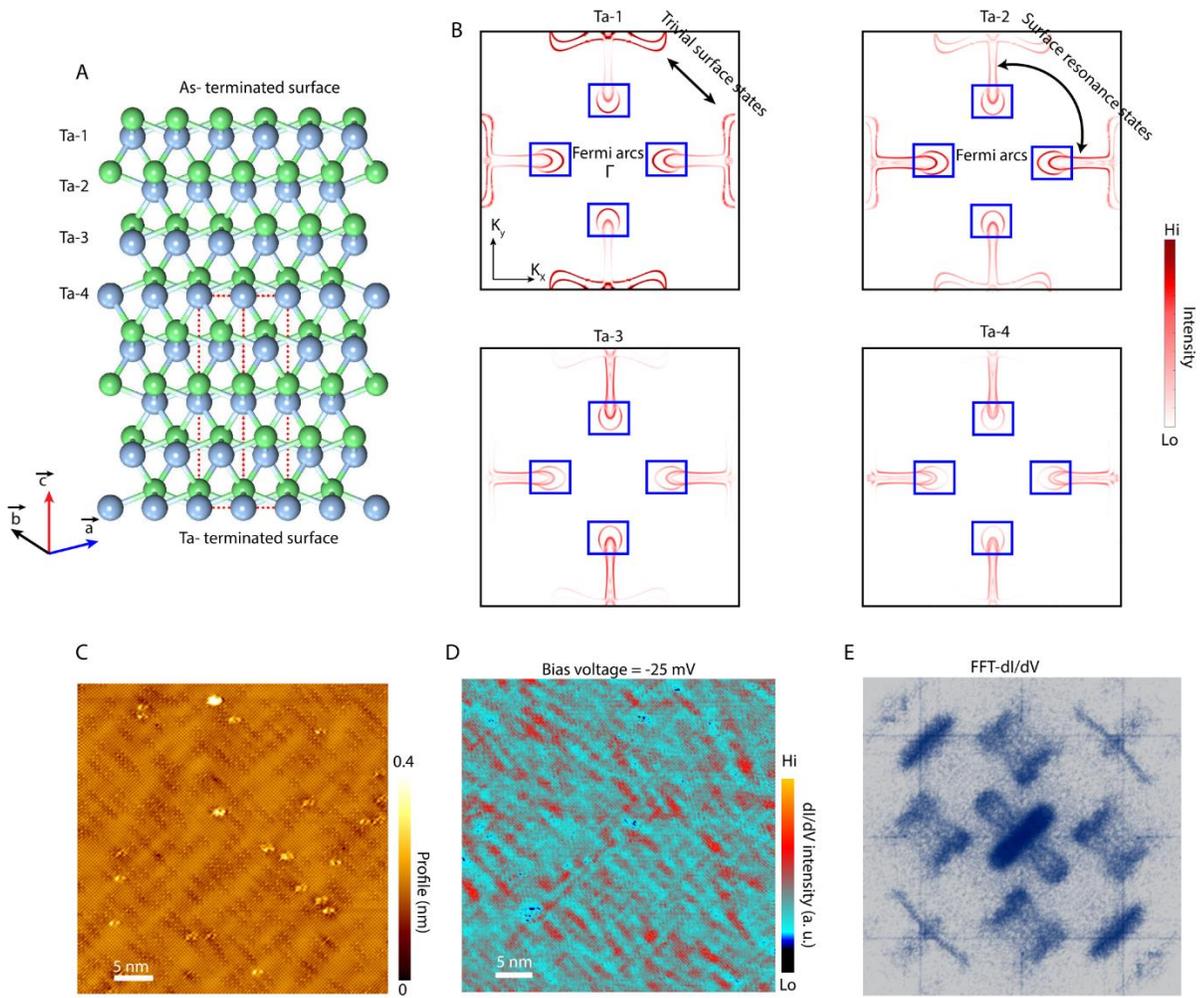

**Fig. S1. Fermi arc states on (001) surface of TaAs.** (**A**) Schematically shows the crystal structure of the TaAs crystal with As-termination on the top (001) face. There are four Ta-layers in a unit cell along $\vec{c}$ direction (Ta-1, Ta-2, Ta-3 and Ta-4), the lattice constant $\vec{c}$ is about 1.2 nm; (**B**) The projection of the calculated surface Fermi contours on each Ta- layer; (**C**) The topographic image of the cleaved (001) face with vacancy defects; (**D**) *dI/dV* conduction map at the region of (C) demonstrates the quasiparticle interference of electrons scattering around the defects; (**E**) Fourier transform of the conduction map (D) presents the scattering wave vectors in the momentum space.

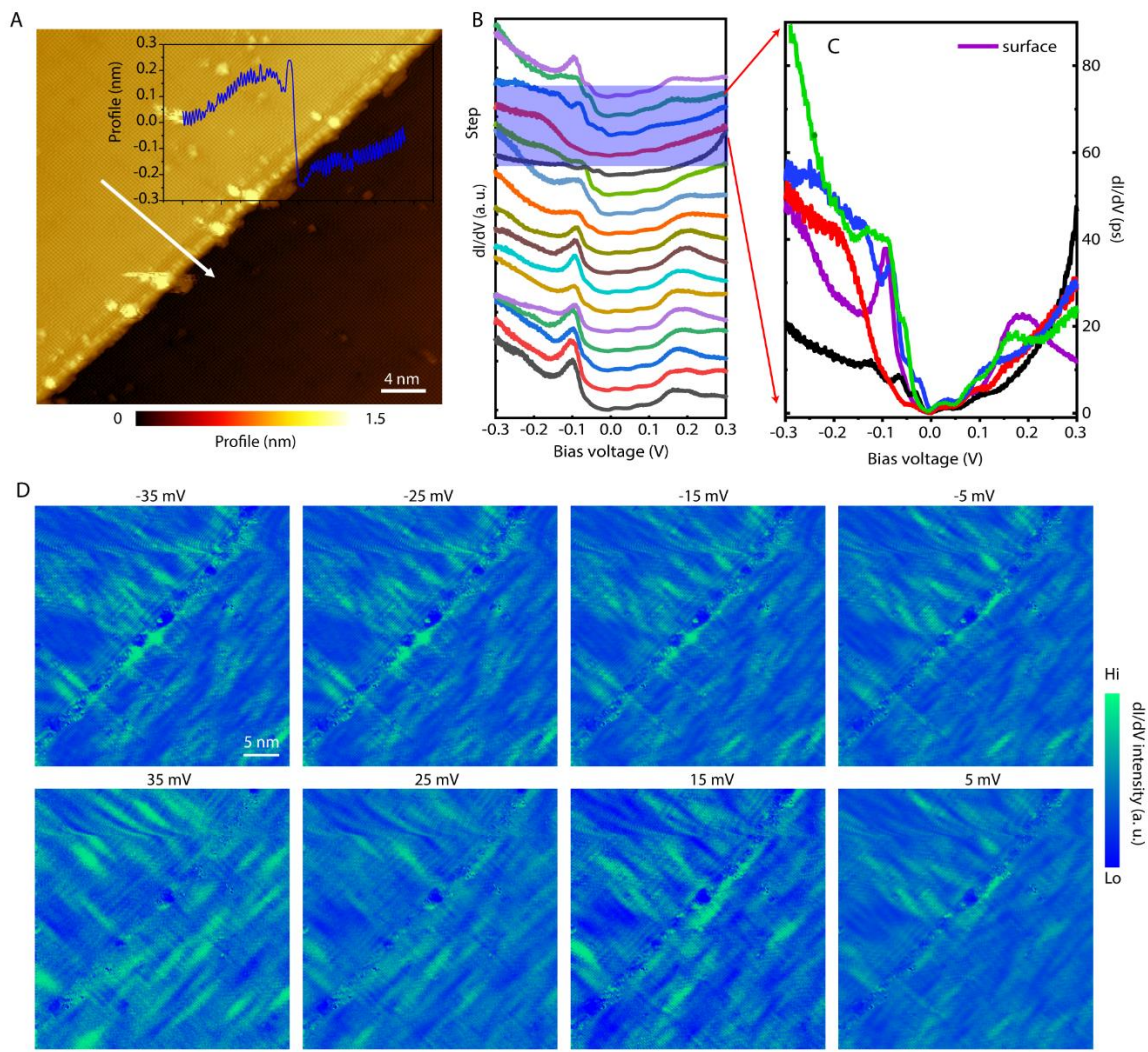

**Fig. S2. Step edge states on (001) surface of TaAs.** (**A**) Topographic image of the (001) face with a step edge. Inset shows the height profile of the step (~0.35 nm); (**B**) The tunneling spectra acquired across the step; (**C**) Near the step edge, the LDOS show position- and energy-dependence where no steadily peaked DOS can be observed; (**D**) *dI/dV* conduction mappings acquired at energy near the Fermi level from the region of (A) show the QPI patterns on the surface, but also no obvious steady and uniform 1D edge states can be observed.

**Surface relaxation in the processed (112) facet**
TaAs is a non-centrosymmetric transition-metal compound with a somewhat complex crystal structure. Cleavage is the method of choice to obtain the desired crystal face although it is particularly challenging for TaAs. In this work, we also develop a procedure with Ar$^+$ bombardment and post-annealing, and obtained the terraced surface as presented in Fig. S3C. However, regardless of nominally the same processing conditions we have observed two types of surface atomic configurations on (112) facet in cleaved and processed samples. Fig. S3A and B show topographic and tunneling current panels acquired simultaneously from the same region of the cleaved (112) surface. The topographic image in Fig. S3A shows a perfect consistency with the Ta- arrangement in the structure model on (112) plane. It indicates that the surface relaxation is negligible during the cleave process. We note however, as revealed by the tunneling signal in Fig. S3B the atomic configurations show certain departure from the structural model. During the measurements, the STM is working in a constant current mode in acquiring a morphological image, there is a loop response time to hold the setpoint, the change of the tunneling current signals can be recorded, simultaneously. The panel in Fig. S3B reveals the modulation of tunneling current signals in each pixel. It contains negligible height information. Tunneling signals contributed from atoms in both the topmost and the second layers can be revealed. The deviation between (A) and (B) implies a strong hybridization of LDOS between the topmost and the underneath atoms. Fig. S3D shows the topographic STM image of the processed (112) surface. Here, the atom configuration is consistent with the tunneling signal image of the cleaved (112) surface in tunneling panel Fig. S3B. It indicates that the annealing process results in the relaxation of the surface atoms to achieve a more stable state.

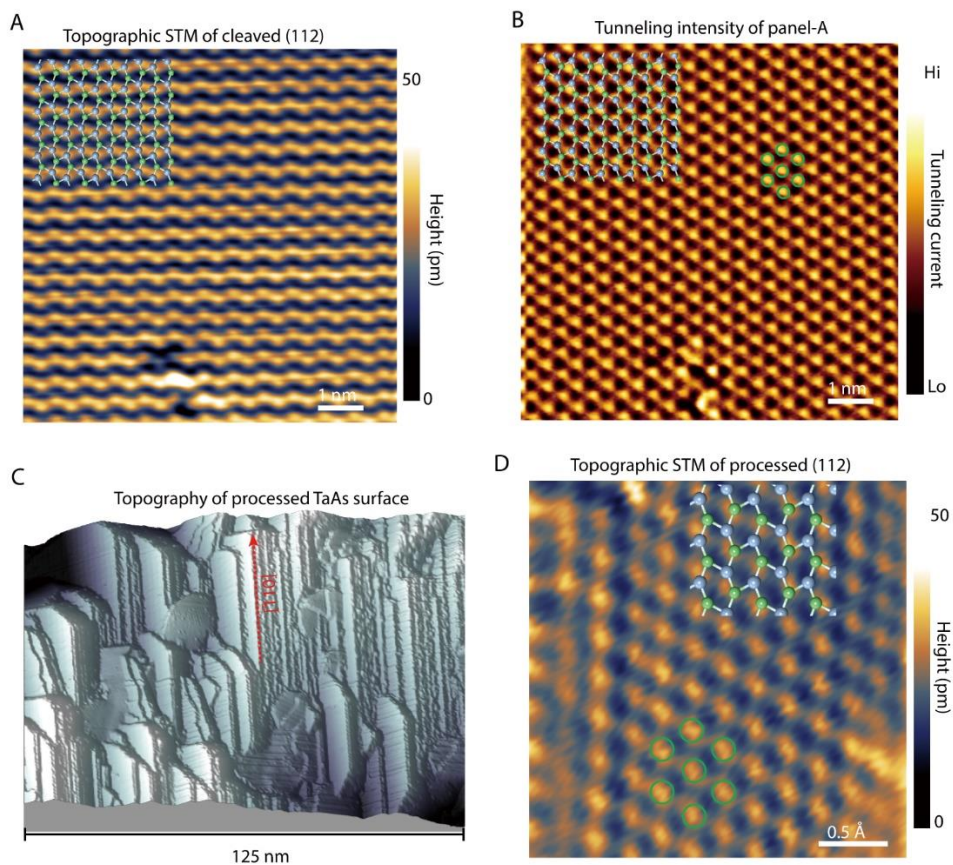

**Fig. S3. Difference between processed and cleaved (112) surfaces.** (**A**) and (**B**) Topographic and tunneling current panels acquired simultaneously from the same region of the cleaved (112) surface; (**C**) The large view topographic image acquired from the processed TaAs sample, where the steps along [$\bar{1}$10] can be observed; (**D**) Atomic resolved STM image acquired from the processed (112) facet.

**Calculations on terraced (110), (112) and (114) surface planes**

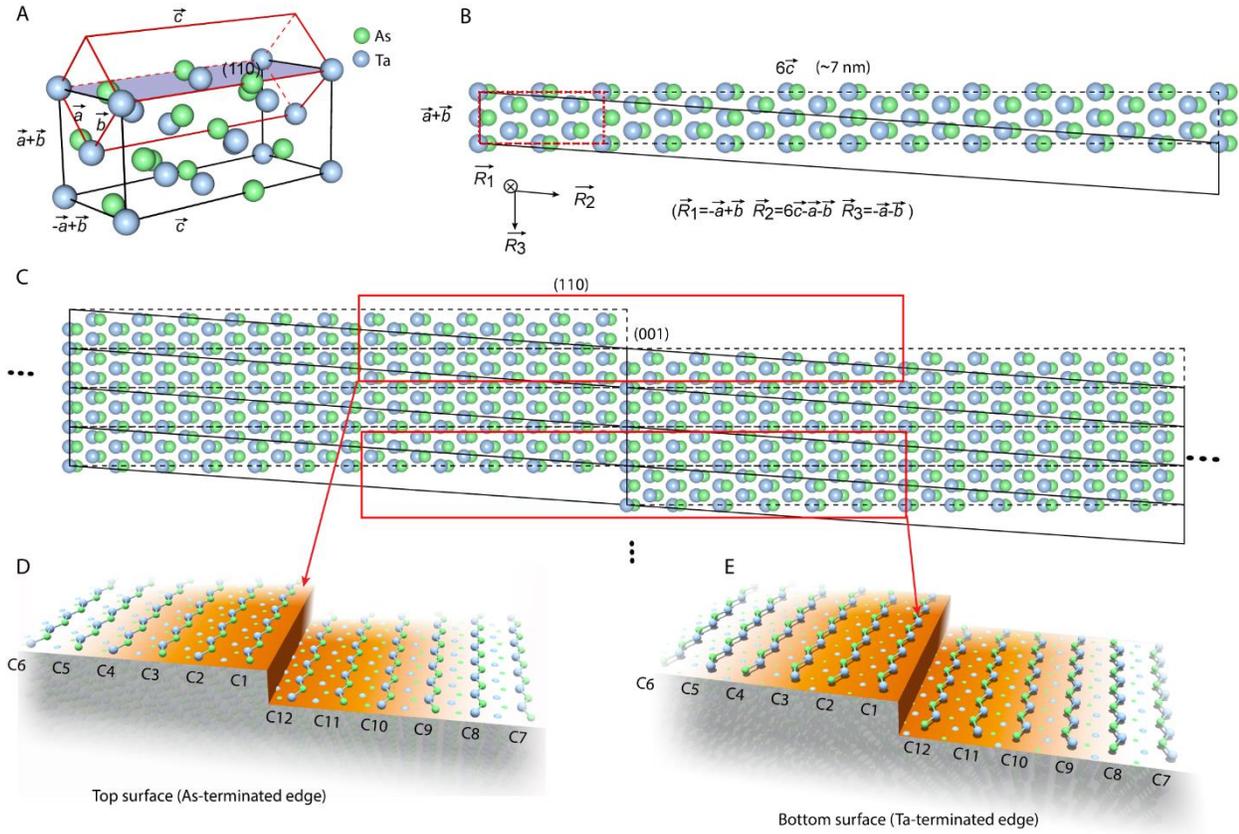

**Fig. S4. Construction of the calculation slab models with (110) terraced planes and atomic-thick (001) exposed at the step ledges.** (**A**) Transformation of TaAs crystal unit cell to a supercell with three lattice vectors $-\vec{a}+\vec{b}$, $\vec{a}+\vec{b}$ and $\vec{c}$, where $\vec{a}$, $\vec{b}$ and $\vec{c}$ are lattice vectors of the TaAs unit cell. The (110) plane is spanned by vectors $-\vec{a}+\vec{b}$ and $\vec{c}$. The black and red boxes represent the supercell and TaAs unit cell, respectively; (**B**) Construction of a larger supercell based on the supercell in (A), it is illustrated by the black solid rhomboid where half of the atoms in the supercell are selected outside the supercell lattice according to the periodic condition. Its lattice vectors are $\vec{R}_1 = -\vec{a}+\vec{b}$, $\vec{R}_2 = 6\vec{c} - \vec{a} - \vec{b}$, $\vec{R}_3 = -\vec{a}-\vec{b}$. $\vec{R}_3$ is the iterative direction, and $\vec{R}_1$, $\vec{R}_2$ define the surface plane $(n,n,2)$ with $n=6$ where the surface states are calculated; (**C**) The schematic illustration of semi-infinite geometry formed in iterative Green's function procedure based on the supercell in (B). Due to the specific atom occupation condition, the periodic step edges jointing (110) and (001) planes are generated automatically. In the calculation slab model, there are two different terraced surfaces (top and bottom), on the top surface (**D**), the step ledges are terminated by As- atoms (constructed by small (001) with As-termination); and on the bottom surface (**E**), the step ledges are terminated by Ta- atoms (constructed by small (001) with Ta-termination). Ta-As chains labeled from C1 to C12 on each (110) terrace can be observed on the top (D) and the bottom (E) surfaces.

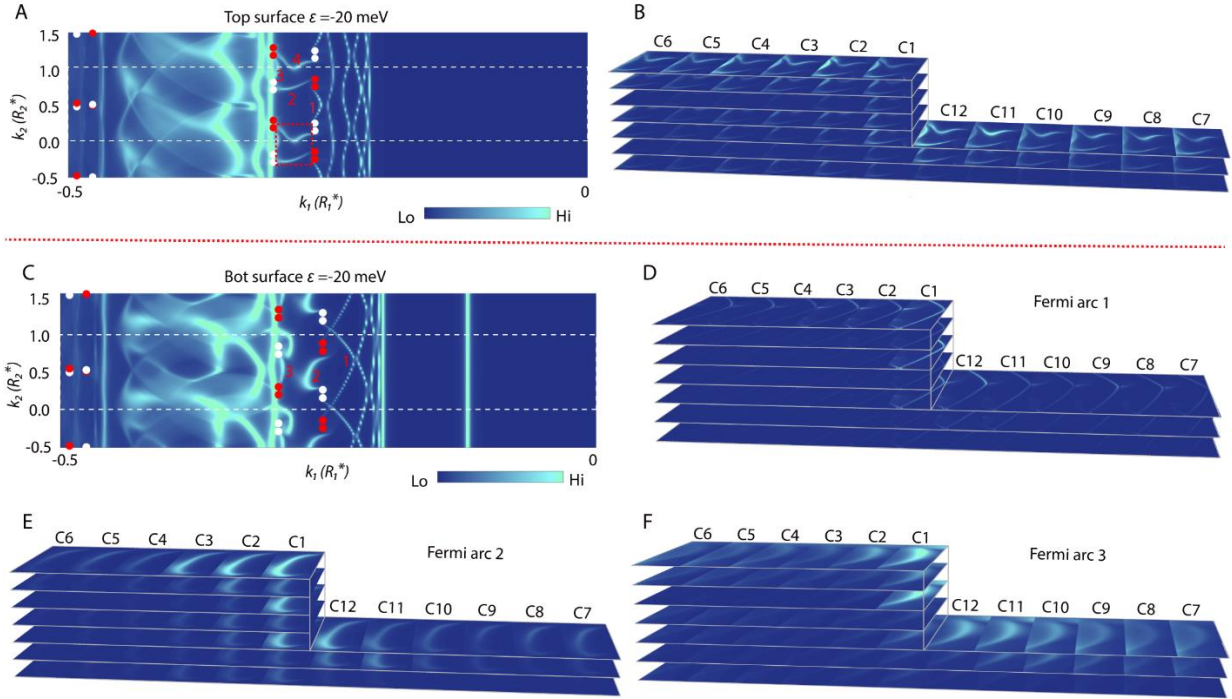

**Fig. S5. All visible Fermi arcs on the top and bottom surface with (110) terraces.** (**A**) Surface FS plots on the surface BZ (half of the whole BZ in the white dashed square) at $\varepsilon = -20$ meV with respect to bulk Fermi energy calculated based on the slab model in Fig. S4D (top surface). FSs on the whole surface BZ can be inferred through time-reversal symmetry. Near the center of the half BZ, there are 4 pairs of chiral Weyl points, and 4 Fermi arcs can be distinguished as numbered from 1 to 4. In the main text arc 1 is presented. Here, other three arcs that projected on the Ta-As chains on (110) terrace have been calculated. Distribution of the projection weight has been shown in (**B**), where the prominent projection weight can all be observed near the step and it decreases steadily as the position moving away from the step, in particular for the direction along (C12 to C7). (**C**) Surface FS plots on the surface BZ (half of the whole BZ in the white dashed square) at $\varepsilon = -20$ meV with respect to bulk Fermi energy calculated based on the slab model in Fig. S4E (bottom surface). There are three arcs that can be identified as numbered as 1, 2 and 3 near the center of the half BZ;(**D**) to (**E**) show the projections of the selected Fermi arc (1, 2 and 3) on the Ta-As chain as numbered in Fig. S4E. The prominent projection weights of the three Fermi arcs are all localized on Ta-As chains near the step edge and decrease steadily as the position far from the step which is in accordance with the calculation results on the top surface and can be well account for the 1D edge states at the step on (110) surface.

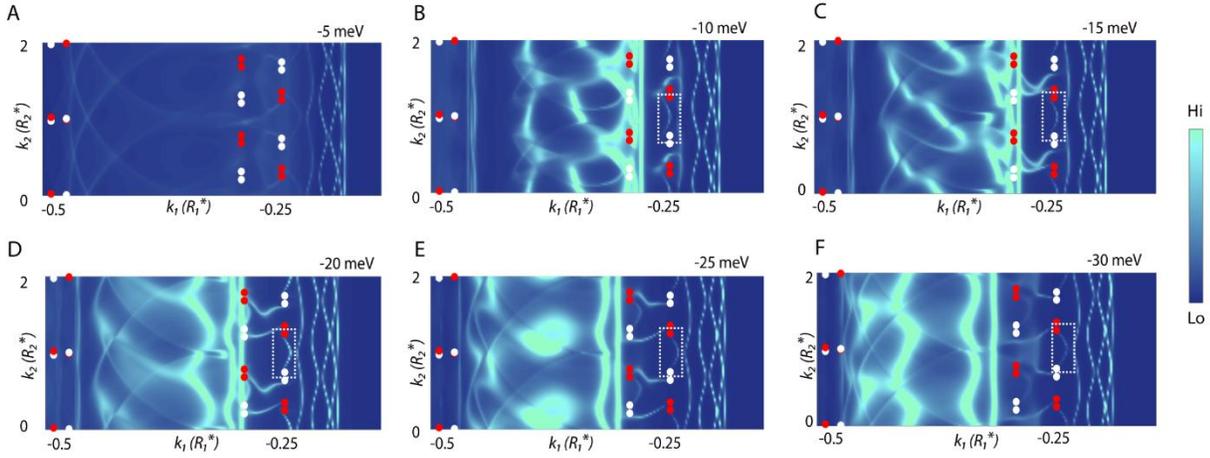

**Fig. S6. Calculation of Fermi arc projection on (110) terraced surface as a function of energy.** (**A**) to (**F**) Surface states on the top surface with (110) terraces at energy -5, -10, -15, -20, -25, -30 meV. The topological Fermi arc states exist near the Fermi level within the energy range between -5 ~ -30 meV.

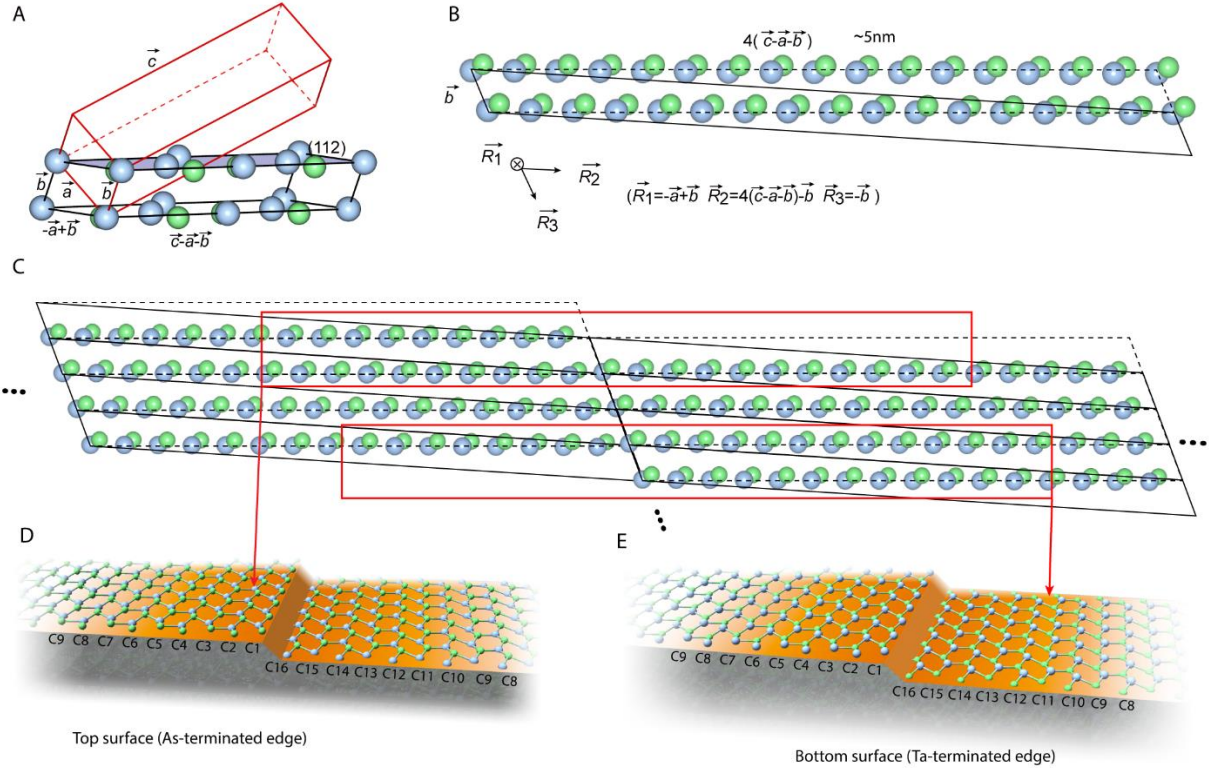

**Fig. S7. Construction of the calculation slab model with (112) terraced planes and atomic-thick (001) exposed at the step ledge.** (**A**) TaAs supercell with the lattice vectors $-\vec{a}+\vec{b}$, $\vec{b}$ and $\vec{c}-\vec{a}-\vec{b}$. The (112) plane is spanned by the supercell lattice vectors $-\vec{a}+\vec{b}$ and $\vec{c}-\vec{a}-\vec{b}$; (**B**) Construction of calculation model for (112) surface planes with periodic steps based on the supercell in (A). The iterative direction $\vec{R}_3$ in the iteration process is $-\vec{b}$. The two vectors $\vec{R}_1=-\vec{a}+\vec{b}$ and $\vec{R}_2=4(\vec{c}-\vec{a}-\vec{b})-\vec{b}$ form the surface plane $(n,n,2n+1)$ with $n=4$ where the periodic step edges jointing (112) and (001) planes are formed automatically and the

surface states are projected on; (**C**) The schematic illustration of semi-infinite geometry formed in iterative Green's function procedure based on the supercell in (**B**). Due to the specific atom occupation condition, the periodic step edges jointing (112) and (001) planes are generated automatically. Two different terraced surfaces are formed meaning the top (**D**) and the bottom (**E**) with the step ledges terminated by As- and Ta- atoms respectively. Each (112) terrace has 16 Ta-As chains on the surface, as labeled from C1 to C16.

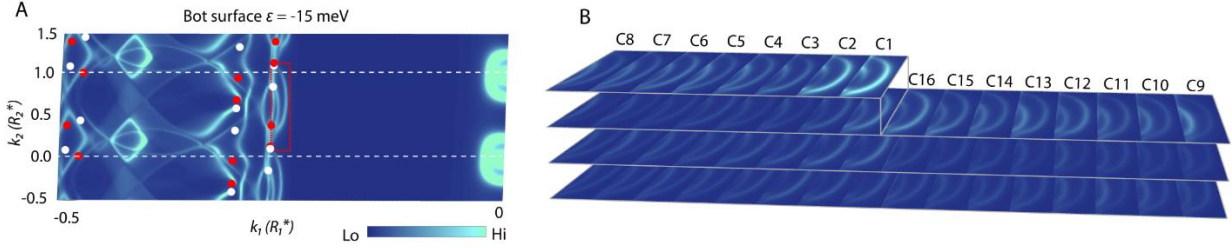

**Fig. S8. Surface states on the bottom surface with (112) terraces.** (**A**) Surface FS plots on half the surface BZ (white dashed square) at $\varepsilon$=-15 meV with respect to bulk Fermi energy calculated based on the slab model in Fig. S7E (bottom surface). The Fermi arc that is clearly isolated from trivial states is selected to calculate the projection weight on Ta-As chains as numbered in Fig. S7E. The projection weight is most prominent at Ta-As chains near the step edge and decreases slowly as the distance from the step increased as shown in (**B**), which is consistent with that on the top surface, as presented in the main text.

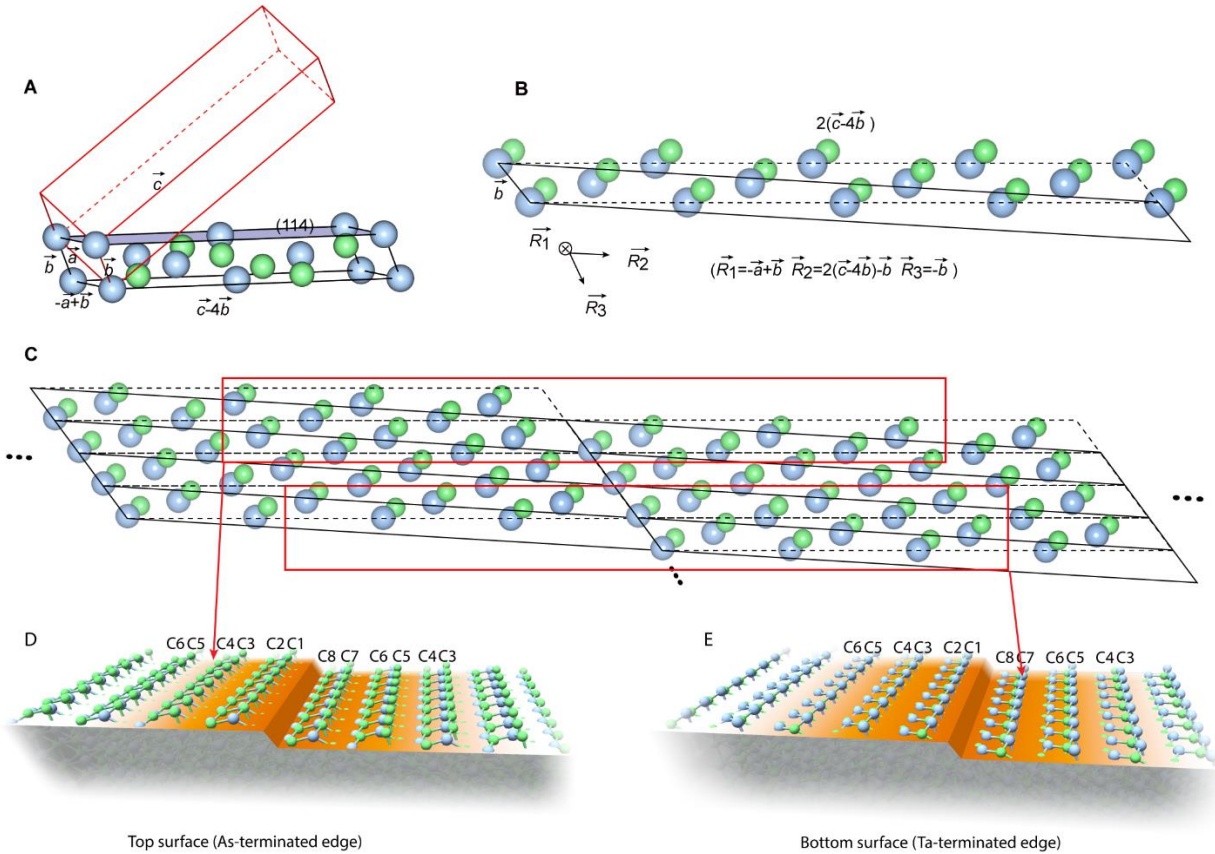

**Fig. S9. Construction of the calculation slab model with (114) terraced planes and atomic-thick (001) exposed at the step ledge.** (**A**) TaAs supercell with the lattice vectors $-\vec{a}+\vec{b}, \vec{b}$ and $\vec{c}-4\vec{b}$; (**B**) Construction of the calculation model for (114) surface. $\vec{R}_1$, $\vec{R}_2$ and $\vec{R}_3$ used in the iterative Green's function scheme are defined as $\vec{R}_1 = -\vec{a}+\vec{b}$, $\vec{R}_2 = 2(\vec{c}-4\vec{b})-\vec{b}$ and $\vec{R}_3 = \vec{b}$. $\vec{R}_3$ is the iterative direction. Here, periodic steps jointing (114) and (001) are formed on the surface plane $(n, n, 4n+1)$ $(n = 2)$ spanned by vectors $\vec{R}_1$ and $\vec{R}_2$; (**C**) The schematic illustration of semi-infinite slab formed in iterative Green's function procedure based on the supercell in (B). Periodic step edges jointing (114) and (001) planes are generated on the (229) surface with flat (114) terrace in the slab with As- terminated edge on the top surface (**D**), and Ta- terminated edge on the bottom surface (**E**). Each terrace has 8 Ta-As chains surface as labeled from C1 to C8.

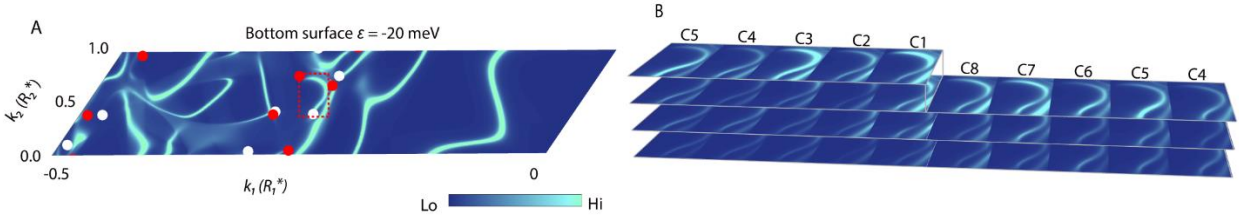

**Fig. S10. 2D Fermi arc surface states on (114) terraced planes.** Surface Fermi surface (FS) plots on (**A**) bottom surface at $\varepsilon$ =-20 meV have been drawn in half the surface BZ. The bottom surface has been described in Fig. S9(E). The Fermi arcs that connecting chiral Weyl points and being isolated from trivial states have been demonstrated; (**B**) show the projection weight distribution of selected Fermi arcs on the bottom surfaces. The arcs can be observed on each of Ta-As chains on the topmost surfaces without significantly enhancing at the step edges on the bottom surfaces, which is consistent with the experimental results that no obvious edge state has been detected. The results confirm there are prominent Fermi arc surface states on (114) crystal facets.